\documentclass[epj]{svjour}

\usepackage{graphics} 
\usepackage{graphicx} 
\usepackage{color}
\usepackage{epsfig}
\usepackage{ulem}

\newcommand{\smatr}{${\cal S}$-matrix}

\begin{document}

\journalname{}

\title{Universal diffusive decay of correlations in gapped one-dimensional systems}
\author{\'A. Rapp\inst{1,2} \and G. Zar\'and\inst{1,3}}

\institute{ Theoretical Physics Department, Institute of Physics, Budapest University of Technology and Economy, Budapest, 1521, Hungary \and Institut f\"ur Theoretische Physik, Universit\"at zu K\"oln, 50937 K\"oln, Germany  \and
Institut f\"ur Theoretische Festk\"orper Physik, Universit\"at Karlsruhe, 76128 Karlsruhe, Germany}

\date{Received 27 June 2008/ Received in final form 29 October 2008\\Published online 24 December 2008   \textcopyright   Sciences, Societ\`{a} Italiana di Fisica, Springer-Verlag 2008}

\abstract{
We apply a semiclassical approach to express finite temperature dynamical correlation functions of gapped spin models analytically. We show that the approach of [\'A. Rapp, G. Zar\'and, Phys. Rev. B \textbf{74}, 014433 (2006)] can also be used for the $S=1$ antiferromagnetic Heisenberg chain, whose lineshape can be measured experimentally. We generalize our calculations to $O(N)$ quantum spin models and the sine-Gordon model in one dimension, and show that in all these models, the finite temperature decay of certain correlation functions is characterized by the same universal semiclassical relaxation function.
\PACS{ {75.10.Pq}{Spin chain models} \and {05.30.-d}{Quantum statistical mechanics} \and  {05.50.+q}{Lattice theory and statistics} }
}

\maketitle 

\section{Introduction}
\label{sec:intro}

The investigation of one-dimensional systems has a long history. In the past decades more and more one-dimensional and quasi-one-dimensional systems have been discovered and investigated experimentally. The discovery of these systems also triggered renewed interest in one-dimensional (1+1 dimensional) models and quantum field theories on the theoretical side. In fact, many of the simplest though non-trivial spin models and quantum field theories are realized in some of these experiments, which thus provides a unique possibility to compare theoretical predictions and reality. One-dimensional models provide the simplest examples of spin liquid systems, can display spin-charge separation, and they can also exhibit quantum phase transitions. Spin chains play a rather special role among one-dimensional systems. In 1931, Bethe solved the $S=1/2$ antiferromagnetic Heisenberg model, and showed that its spectrum is gapless~\cite{Bethe}. His method has been generalized, and is now widely used to study integrable one-dimensional systems. More than 20 years ago, however, Haldane pointed out that in contrast to the spin $S=1/2$ Heisenberg model, quantum Heisenberg chains with an integer spin do have a gap in the excitation spectrum~\cite{Haldane}. This astonishing difference between the half-integer and integer spin cases is by now accepted and well understood analytically~\cite{Affleck} and numerically~\cite{Deisz}, and the existence of the gap was also confirmed experimentally~\cite{haldane-materials-1,haldane-materials-2,haldane-materials-7,haldane-materials-3,haldane-materials-4,haldane-materials-5,haldane-materials-6}. Although zero temperature integer-spin chains and their static properties at finite temperatures have been studied extensively, understanding their finite temperature dynamical behavior still poses a major theoretical challenge. Not even the dynamics of the $S=1$ Heisenberg antiferromagnet has been fully understood so far. Characterizing these dynamical properties in detail would be, however, crucial in order to interpret inelastic neutron scattering~\cite{haldane-materials-1,haldane-materials-2,haldane-materials-7,haldane-materials-3,haldane-materials-4} or nuclear magnetic resonance~\cite{haldane-materials-5,afm-ladder-1} experiments. One of our goals in this paper shall be to describe the dynamical correlations of the $S=1$ antiferromagnetic Heisenberg chain, which are related to the inelastic lineshape measured directly via neutron scattering by a Fourier transform. So far two methods proved to be efficient to study the dynamical correlations at finite temperature:

(A) a field theoretical approach based upon a form factor expansion~\cite{formfactor1,formfactor2,formfactor3,formfactor4,Konik-nuclphys} of the dynamic structure factor (DSF). This method suggests that the lineshape of the dynamical susceptibility of the $S=1$ antiferromagnetic Heisenberg chain around $q=\pi$ is rather asymmetric around the gap, and this asymmetry grows with increasing temperature~\cite{Konik-lineshape}. 

(B) in an approach due to Sachdev and Young, the quasiparticles are described semiclassically and only their collisions are treated quantum mechanically~\cite{semiclass-SachdevYoung}. 

\begin{figure}
\centering
\resizebox{0.5\columnwidth}{!}{\input{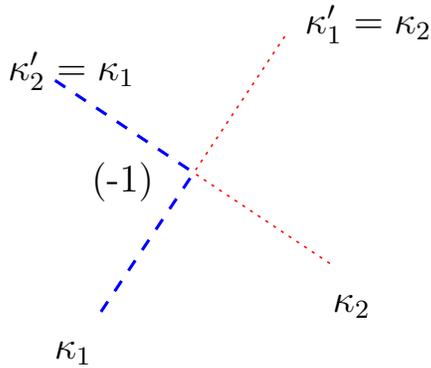}}
\caption{ (Color online) The reflective \smatr. During the semiclassical evolution of the system, the sequence of the internal quantum numbers of the elementary excitations remains invariant. }
\label{fig:refl_smatrix}
\end{figure}

In this paper we shall use the second, semiclassical method to calculate the finite temperature dynamical correlation function analytically for various spin models. This approach is based on the observation that in the spin models investigated here, quasiparticles have a dispersion relation $\epsilon_k$ with a spectral gap $\Delta$. Since the quasiparticle density, $\rho \sim e^{-\Delta/T}$, is exponentially small at low temperatures, $T \ll \Delta$, the thermal de~Broglie wavelength, $\lambda_{\rm dB} \sim 1/\sqrt{T}$, is negligible compared to the average separation of the quasiparticles, $\sim \rho^{-1}$. Hence the quasiparticles behave quasiclassically except for the unavoidable collisions in $d=1$ dimension. Generic gapped systems have quasiparticles which posses some internal quantum number $\kappa$. For a number of models~\cite{rapp-potts,semiclass-SachdevYoung,sinegordon,semiclass-DamleSachdev,ON-smatrix}, it has been verified that at low temperatures, in the long wavelength limit the \smatr~describing quasiparticle collisions becomes universally reflective (see Fig.\ref{fig:refl_smatrix}),
\begin{equation}
	{\cal S}_{\kappa_1 \kappa_2}^{\kappa_1' \kappa_2'} \to (-1) \delta_{\kappa_1}^{\kappa_2'}  \delta_{\kappa_2}^{\kappa_1'} \;.
\end{equation}
We shall use this simple structure of the \smatr~to compute certain time-dependent correlation functions. In particular, we shall evaluate correlation functions of the form
\begin{equation}
	C(x,t) = \langle \hat O(x,t) \hat O(0,0) \rangle\;. \label{eq:def:corrfunc}
\end{equation}
Here the operator $\hat O(x,t)$ depends on the particular model studied: it corresponds to the operators $\hat S^z(x,t)$ in the antiferromagnetic Heisenberg chain and $\hat n^z(x,t)$ in the O(3) quantum rotor model, and it will be a vertex operator in the sine-Gordon model.

Although the correlation function in equation~(\ref{eq:def:corrfunc}) has already been computed for the sine-Gordon model by Damle and Sachdev in reference~\cite{sinegordon} analytically, they determined the $n^z -n^z$ correlation function for the $O(3)$ rotor model only numerically in their previous work, reference~\cite{semiclass-DamleSachdev}. Here we show in detail that this correlation function can be obtained also analytically for all $O(N)$ models using the method of reference~\cite{rapp-potts}. As we shall see, for the models considered here, the semiclassical relaxation function assumes the same universal form, as already stated in references~\cite{rapp-potts} and \cite{semiclass-DamleSachdev}.

\begin{figure}
	\centering
	\resizebox{0.95\columnwidth}{!}{\input{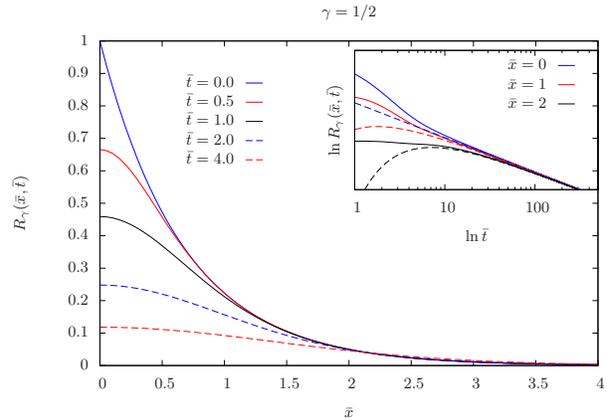}}
	\caption{ (Color online) Relaxation function $R_\gamma(\bar x,\bar t)$ for $\gamma=1/2$. The inset shows the long time behavior $\bar t \gg 1$, the dashed lines correspond to the asymptotic diffusive form in equation~(\ref{eq:diff_form}). \label{fig:Q3}} 
\end{figure}

\begin{figure}{t}
	\centering
	\resizebox{0.95\columnwidth}{!}{\input{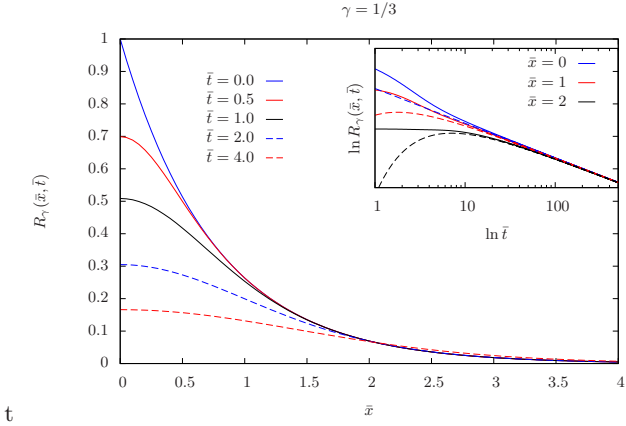}}
	\caption{ (Color online) Relaxation function $R_\gamma(\bar x,\bar t)$ for $\gamma=1/3$. The inset shows the long time behavior $\bar t \gg 1$, the dashed lines correspond to the asymptotic diffusive form in equation~(\ref{eq:diff_form}). \label{fig:Q4}} 
\end{figure}

In many cases, the operator $\hat O$ creates (or destroys) a quasiparticle. Therefore, at finite temperatures, we factorize the correlation function as
\begin{equation}
	C(x,t) = C(x,t)_{T=0} R_\gamma(\bar x,\bar t), \label{eq:def:relax}
\end{equation}
where $C(x,t)_{T=0}$ corresponds to the coherent propagation of a quasiparticle, while the relaxation function $R_\gamma(\bar x,\bar t)$, defined through equation~(\ref{eq:def:relax}), describes the scattering with the thermally excited quasiparticles. For all models considered here in the semiclassical limit, the relaxation function is given by
\begin{eqnarray}
R_\gamma(\bar x ,\bar t) &=& \int\limits_{-\pi}^{\pi} \frac{d \phi}{2\pi} 
\frac{(1-\gamma^2)  \cos{ \left( \sin({\phi})  \bar{x} \right) }}{ \gamma^2+2\gamma\cos{\phi}+1}  \label{eq:relaxfv} \\  
&&  {\rm exp}\left[ - | \bar{t} |  (1-\cos{\phi}) \left( \frac{1}{\sqrt{\pi}} e^{-u^2} + u {\rm erf}(u) \right) \right]  \nonumber \;, 
\end{eqnarray}
where $\gamma$ is a model-specific parameter, and $u=\bar x / \bar t$, $\bar x= x / \xi_c$ and $\bar t = t / \tau_c$ are the dimensionless velocity, position and time. Here the characteristic length and time scales are defined as $\xi_c = \gamma \; \sqrt{\frac{2\pi \; c^2}{T\Delta}} e^{\Delta/T}$, and $\tau_c = \gamma \; \frac{\sqrt{\pi}}{T} e^{\Delta/T}$, respectively, with $\Delta = \epsilon(k)\vert_{k\to 0}$ the quasiparticle gap and $c^2 = \Delta [d^2 \epsilon_k / dk^2]\vert_{k \to 0}$ the square of the characteristic ``velocity of light''. The relaxation function is shown in Fig.~\ref{fig:Q3} and Fig.~\ref{fig:Q4}, for the parameters $\gamma=1/2$ and $\gamma=1/3$, respectively. In the $Q$-state quantum Potts model, $\gamma = 1/(Q-1)$~\cite{rapp-potts}. We shall prove that $\gamma = 1/N$ for the $O(N)$ quantum rotor models, while in the sine-Gordon model we find $\gamma = -{\rm cos} \left(\frac{2\pi \eta}{\gamma'}\right)$, with $\gamma'$ and $\eta$ defined later~\cite{sinegordon}. Note that in the case of the sine-Gordon model, $\xi_c = \frac{1}{2} \; \sqrt{\frac{2\pi \; c^2}{T\Delta}} e^{\Delta/T}$, and $\tau_c = \frac{1}{2} \; \frac{\sqrt{\pi}}{T} e^{\Delta/T}$, respectively. The relaxation function for $\gamma < 1$ is diffusive for long time scales $1 \ll \bar t$,
\begin{equation}
R_\gamma(\bar x,\bar t) \sim \frac{1}{\sqrt{4 \pi \bar D \bar t}} e^{ - \frac{\bar x ^2}{4 \bar D \bar t}} \;, \label{eq:diff_form}
\end{equation}
where $\bar D = \frac{1}{2\sqrt{\pi}}$ is the dimensionless diffusion constant. On the other hand, in the transverse field Ising model, $\gamma = 1 $ ($Q=2$). In this special limit, $\gamma \to 1$, the relaxation becomes exponential instead of being diffusive~\cite{semiclass-SachdevYoung}. Exponential behavior also appears under the assumption of a transmissive/non-interacting scattering matrix $ {\cal S}_{\kappa_1 \kappa_2}^{\kappa_1' \kappa_2'} \to (-1) \delta_{\kappa_1}^{\kappa_1'} \delta_{\kappa_2}^{\kappa_2'} $~\cite{rapp-potts}. We believe, however, that generic lattice spin models always have 
a reflective scattering matrix, and special conditions and cut-off schemes are required to obtain a 
diagonal \smatr.

This paper is organized as follows. We shall discuss the  semiclassical correlations of $S=1$ antiferromagnetic Heisenberg model and their relation to the correlation functions in the $O(3)$ quantum rotor model in Sec.~\ref{sec:heis}. These results will be generalized to the $O(N)$ rotor chains in Sec.~\ref{sec:ON-rotor}. 
In Sec.~\ref{sec:sGordon}, we derive using our approach the semiclassical relaxation function of the sine-Gordon model, first derived in a somewhat different way and form by Damle and Sachdev~\cite{sinegordon}. The conclusions are presented in Sec.~\ref{sec:concl}.

\section{The AF Heisenberg chain and the quantum rotor model}
\label{sec:heis}

The $S=1$ Heisenberg chain is defined by the Hamiltonian
\begin{equation}
\hat H ^{\rm Heis} = J \sum_i \hat \mathbf{S}_i \hat \mathbf{S}_{i+1}   \;,\label{eq:Heis_chain}
\end{equation}
where $\hat\mathbf{S}_i$ is an $S=1$ spin at site $i$, and $J>0$ is the antiferromagnetic exchange coupling. As first proved by Haldane, antiferromagnetic spin waves have a gap, if $S$ is an integer~\cite{Haldane}. Haldane's prediction is in good agreement with many experiments on different quasi-1D materials~\cite{haldane-materials-1,haldane-materials-2,haldane-materials-7,haldane-materials-3,haldane-materials-6,haldane-materials-5,haldane-materials-4}, although the SU(2) symmetrical Heisenberg model provides usually only an approximate description of these experimental systems, since perturbations like magnetic anisotropy, lattice distortions, disorder, etc., are always present in physical systems.

Low-energy excitations of the integer spin Heisenberg model in equation~(\ref{eq:Heis_chain}) can be mapped onto the $O(3)$ rotor chain~\cite{Haldane,Affleck}, defined by the following Hamiltonian,
\begin{equation}
\hat H ^{\rm rotor}= \frac{\tilde J g}{2} \sum_i \hat \mathbf{L}_i^2 - \tilde J \sum_i \hat \mathbf{n} _i \hat \mathbf{n} _{i+1}\;. \label{eq:rotor}
\end{equation}
Here $\hat \mathbf{n} _i$ is the position operator (orientation) of the rotor on site $i$ with the constraint $\hat \mathbf{n}_i^2 = 1$, and $\hat \mathbf{L}_i = \hat \mathbf{n}_i \times \hat \mathbf{p}_i$ is its angular momentum operator satisfying the commutation relations $[ \hat L_i^\alpha, \hat L_i^\beta] = i\epsilon_{\alpha\beta\gamma}\hat L_i^{\gamma}$ and $[\hat L_i^\alpha, \hat n_i^\beta] = i\epsilon_{\alpha\beta\gamma}\hat n_i^{\gamma} $. The mapping can be done most straightforwardly in the path integral formalism by representing spins in terms of coherent states $\mathbf{N}_i(\tau) \Leftrightarrow \hat \mathbf{S}_i(\tau)$. The unit modulus field $\mathbf{N}_i(\tau)$ can then be parametrized~\cite{Affleck,Haldane,Sachdev_book} by slowly varying fields as
\begin{equation}
\mathbf{N}_i(\tau) = (-1)^i \mathbf{n}_i(\tau) \sqrt{1 - a^2 \mathbf{L}^2_i(\tau)} + a \mathbf{L}_i(\tau) \;, \label{eq:def:vector2}
\end{equation}
where $\mathbf{n}$ describes the staggered and $\mathbf{L}$ the uniform component of the Heisenberg spins, and $a$ is the lattice spacing. In the $S=1$ case, the excitations corresponding to both fields $\mathbf n (\tau)$ and $\mathbf L (\tau)$ are massive~\cite{Haldane}. Integration over $\mathbf L$ leads to the non-linear $\sigma$-model, defined by the action
\begin{equation}
{\cal A}^{{\rm nl}\sigma{\rm m}} = \frac{3}{2\tilde c \tilde g} \int\limits_0^{\beta} d\tau\int dx [(\partial_\tau \tilde {\mathbf n}(x,\tau) )^2  +\tilde c^2 ( \partial_x \tilde {\mathbf n}(x,\tau) )^2 ] \;,
\end{equation}
where $\tilde g$ is a coupling constant, $\tilde c$ has a dimension of velocity, and the field $\tilde {\mathbf n}(x,\tau)$ satisfies the constraint $\tilde {\mathbf n}^2(x,\tau) = 1$. The non-linear $\sigma$-model is the continuum theory of the rotor chain, equation~(\ref{eq:rotor})~\cite{Sachdev_book}.

Let us now discuss the properties of the rotor model defined by equation~(\ref{eq:rotor}). For $g \gg 1$, in the ground state, all rotors need to be in the $L=0$ state to minimize kinetic energy. The lowest energy excitations form a triplet with quantum numbers $L^z \equiv \lambda = -1,0,1$. In $d=1$ dimension, the qualitative structure of the low-energy spectrum is the same for any $g>0$ and the excitations have a gap $\Delta (g)$~\cite{semiclass-DamleSachdev,Sachdev_book}. At low enough temperatures, $T \ll \Delta$, the gap allows us to apply a semiclassical approximation. 

\begin{figure}
\resizebox{0.95\columnwidth}{!}{\input{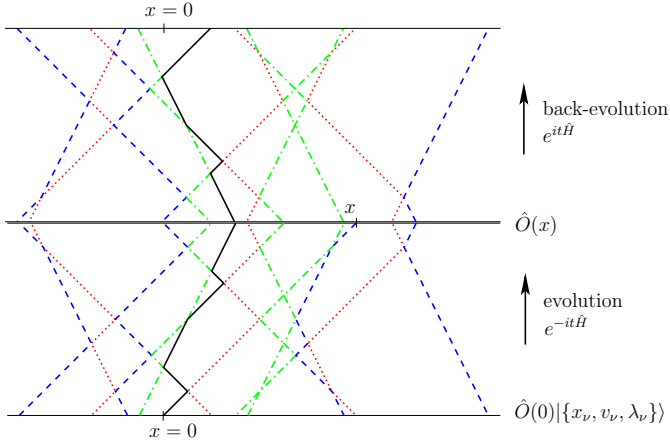}}
\caption{ \label{fig:semiclass_dynamics} (Color online) Semiclassical dynamics in the $O(3)$ rotor model. Due to the reflective structure of the \smatr, the \textit{sequence} of the quantum numbers $\lambda_\nu$ does not change in time. }
\end{figure}

Let us focus on the dynamical correlation function of the rotor model, and follow the same steps as for the quantum Potts model~\cite{rapp-potts}. In the semiclassical limit, the correlation function can be approximated as:
\begin{eqnarray}
C^{\rm rotor}(x,t) &=& \langle \hat n^z(x,t) \hat n^z(0,0) \rangle_{\hat H^{\rm rotor}}  \label{eq:corrfunc_0} \\
&\approx& \sum_{\{\lambda_\nu\}} \int \prod_\nu dx_\nu \prod_\nu dv_\nu  \Big{[} P (\{ x_{\nu},v_{\nu},\lambda_{\nu} \}) \nonumber \\
&& \times \langle  \{x_{\nu},v_{\nu},\lambda_{\nu} \}| \hat n^z(x,t) \hat n^z(0,0) |\{x_{\nu},v_{\nu},\lambda_{\nu} \}\rangle \Big{]} \;, \nonumber
\end{eqnarray}
where the function $P(\{x_\nu , v_\nu ,  \lambda_\nu \})$ is the probability density of having quasiparticles with velocities $v_\nu$ and internal quantum numbers $L^z_\nu = \lambda_\nu (\lambda_\nu=\pm1,0)$ at positions $x_\nu$. In the low density limit, $P(\{x_\nu , v_\nu ,  \lambda_\nu \})$ factorizes as
\begin{equation}
P(\{x_\nu , v_\nu , \lambda_\nu \}) = \frac{1}{L^M} \frac{1}{3^M} \prod_{\nu} P(v_\nu) \;, \label{eq:def:semiclass_distr}
\end{equation}
with the distribution of the velocities given by the Maxwell-Boltzmann statistics, 
\begin{equation}
P(v) = \sqrt{\frac{ \Delta}{ 2\pi c^2 T}} {\rm exp}\left( - \frac{\Delta v^2}{2 c^2 T} \right) \;. \label{eq:Boltzmann}
\end{equation}
Calculating the matrix elements for $\hat n^z (\Leftrightarrow \cos \theta)$ with the first few spherical harmonics, one finds that $\hat n^z$ either creates a quasiparticle ($L = 1$) with $L^z=\lambda=0$ at $x=0$ with some velocity $v$ or destroys one already present in the configuration $\{x_{\nu},v_{\nu},\lambda_{\nu} \}$. The probability of latter is exponentially small since at $T \ll \Delta$ the quasiparticle density $\rho$ is low, so we shall neglect this. Due to the collisions with the thermally excited particles, there are only certain configurations where the quantummechanical overlap in equation~(\ref{eq:corrfunc_0}) will be non-zero. Similar to the Potts model discussed in reference~\cite{rapp-potts}, the $O(3)$ rotor model has a purely reflective scattering matrix~\cite{semiclass-DamleSachdev,ON-smatrix},
\begin{equation}
{\cal S}^{\lambda_1' \lambda_2'}_{\lambda_1 \lambda_2} = (-1) \delta_{\lambda_1}^{\lambda_2'} \; \delta_{\lambda_2}^{\lambda_1'} \;. \label{eq:smatr-rotor}
\end{equation}
This structure implies that the \textit{sequence} of the quantum numbers $\lambda_\nu$ does not change in time (see Fig.~\ref{fig:semiclass_dynamics}). Thus if the extra particle created by $\hat n^z$ collides with $N_+$ particles from the right and $N_-$ particles from the left, then to have a non-vanishing overlap, the first $n=N_+-N_-$ particles to the right must also have $\lambda_{\nu}=0$. At time $t$ a quasiparticle must be destroyed by $\hat n^z(x,t)$ such that the overlap in equation~(\ref{eq:corrfunc_0}) does not vanish. These observations are analogous to the arguments given for the paramagnetic side of the quantum Potts model~\cite{rapp-potts}. Taking also the phase factors into account, we get an expression similar to that obtained for the paramagnetic side of the quantum Potts model
\begin{equation}
C^{\rm rotor}(x,t)=C_{T=0}^{\rm rotor} (x,t) \tilde{R}(x,t) \;,   \label{eq:corrfunc_1}
\end{equation}
where the first term is the average of a phase factor coming from the propagation of a single particle from $(0,0)$ to $(x,t)$~\cite{semiclass-DamleSachdev}
\begin{equation}
	C_{T=0}^{\rm rotor} (x,t) = \frac{{\cal Z}}{2\pi} K_0(\Delta \sqrt{x^2/c^2 -t^2})\;, \label{eq:propagator}
\end{equation}
where $K_0$ is the modified Bessel function of the second kind and $\cal Z$ is a non-universal quasiparticle residue. Well within the light cone, $x \ll ct$, one can recover the Feynman propagator of a particle with a mass of $\Delta$:
\begin{equation}
C_{T=0}^{\rm rotor} (x \ll ct,t) \sim   e^{-i\Delta t} \sqrt{\frac{\Delta}{2\pi i t}} {\rm exp} \left(i\frac{\Delta x^2}{2 c^2 t} \right) \;. 
\end{equation}
The second term in equation~(\ref{eq:corrfunc_1}) is given by
\begin{eqnarray}
\tilde{R}(x,t) &=& \sum_{n=-\infty}^\infty {(-1)^n \over 3^{|n|}}
\left\langle \delta_{n,\sum_\nu [\Theta(x - x_\nu - v_\nu t) -  \Theta(-x_\nu)]}
\right\rangle_{\{v_\nu,x_\nu\}} \nonumber \\
&=& R_{1/3}(x/\xi_c,t/ \tau_c). \label{eq:relax0}
\end{eqnarray}
In the middle expression, we already took the average over the angular momentum components $\{ \lambda_\nu  \}$, leading to the factor $1/3^{|n|}$. The expression in the average is the probability that the excited particle drifts to the right by $n$. The factor of $(-1)^n$ corresponds to the net phase accumulated in course of the scattering with the quasiparticles. The average over the velocities and coordinates can be carried out analytically to obtain the result on the right hand side of equation~(\ref{eq:relax0})~\cite{rapp-potts}.

Having derived the semiclassical correlation function of the O(3) rotor model, the correlation function of the antiferromagnetic Heisenberg chain
\begin{equation}
 C^{zz}(x,t) \equiv \langle  \hat S^z(x,t) \hat S^z(0,0)  \rangle \;,
\end{equation}
can be approximated at low temperatures as follows. As it can be seen from equation~(\ref{eq:def:vector2}), and proved more rigorously in reference~\cite{Sachdev_book}, while the fluctuations around $q=0$ in the Heisenberg chain are described by the $L^z-L^z$ correlation function of the $O(3)$ quantum rotor chain,~\cite{Sachdev_book} the dynamics at $q=\pi/a$ is related to the $n^z-n^z$ correlation function. As a consequence, in the low temperature limit, the dynamic structure factors of the Heisenberg and rotor models are simply related as
\begin{equation}
S^{zz} (q = \pi/a + k, \omega ) \sim S^{\rm rotor}(k,\omega) \;.
\end{equation}
The correlation function that we calculated for the rotor model analytically, has been obtained by Damle and Sachdev numerically using a stochastic sampling. The $L^z-L^z$ correlation function has been evaluated analytically in reference~\cite{semiclass-DamleSachdev}.

 \section{The $O(N)$ rotor chain}
\label{sec:ON-rotor}

As a generalization of the $O(3)$ quantum rotor chain, let us now investigate the correlations of rotor models with $O(N)$ symmetry, defined by the Hamiltonian
\begin{equation}
\hat H ^{O(N)}= \frac{J g}{2} \sum_i \hat \mathbf{L}_i^2 - J \sum_i \hat \mathbf{n} _i \hat \mathbf{n} _{i+1}\;, \label{eq:ON-rotor}
\end{equation}
where $\hat \mathbf{n}_i$ is the $N \geq 3$-dimensional position operator of the rotor at site $i$ with the constraint $\hat \mathbf{n}^2_i = 1$. The angular momentum operator components are given by $\hat L_{\alpha\beta} = \hat n^\alpha \hat p_\beta -\hat n^\beta \hat p_\alpha$, where $[\hat n^\alpha, \hat p_\beta ]= i \delta_{\alpha\beta}$, and the square of the angular momentum operator is defined as
\begin{equation}
	\hat \mathbf{L}^2 = \frac{1}{2} \sum_{\alpha\neq\beta} \hat L_{\alpha\beta}^2 \;.
\end{equation}
It is easy to see that the operator $n^\alpha$ creates a member of an $N$-component quasiparticle 
multiplet. The reasoning is based on the Wigner-Eckart theorem: The ground state is 
structureless, and transforms as an $O(N)$ singlet under $O(N)$ rotations. However, from the commutation relations it trivially follows that $\hat n^\alpha$-s form the components of an $N$-dimensional irreducible tensor operator. As a consequence, to have a non-vanishing matrix element, $\langle \beta \vert \hat n^\alpha \vert 0 \rangle$, the state $\beta$ must transform according to the same irreducible representation as 
$\hat n^{\alpha}$, moreover, the matrix elements must be proportional to $\delta_{\alpha,\beta}$. This implies that single particle excitations of momentum $k$ and having a finite overlap with  $\hat n^\alpha\vert 0 \rangle $ must form $N$-fold degenerate multiplets, $\vert k,\alpha \rangle (\alpha=1,..,N)$. Moreover, since they transform as the components of $\hat n^\alpha$, they must also be orthogonal to each other, and have the matrix elements $\langle k,\beta \vert \hat n_\alpha\vert 0 \rangle = A(k) \delta_{\alpha,\beta}$, with $A(k)$ a momentum-dependent form factor.

Similar to the $O(3)$ rotor model, the continuum limit of the $O(N)$ rotor chain maps onto the 1+1-dimensional $O(N)$ sigma model.  The asymptotic \smatr~of the continuum $O(N)$ non-linear $\sigma$-model is purely reflective, and here we shall assume that this holds also for the lattice-regularized Hamiltonian~\cite{ON-smatrix}. 

It is thus straightforward to generalize our discussion of the $O(3)$ rotor model to this case and 
we find that, in the semiclassical limit, the correlation functions of the $O(N)$ rotor chains can also be factorized as
\begin{eqnarray}
	C^{O(N)}(x,t) &=& \langle \hat n^N(x,t)  \hat n^N(0,0) \rangle \nonumber \\
	&=& C_{T=0}^{O(N)}(x,t) \; R_{\gamma=1/N}(\bar x, \bar t),
\end{eqnarray}
where $C_{T=0}^{O(N)}(x,t)$ corresponds to the $T=0$ temperature propagation of a quasiparticle. Hence the universal relaxation function $R_{\gamma}(\bar x,\bar t)$ describes the decay of correlations of $O(N)$ rotor and non-linear sigma models in the semiclassical limit.

\section{The sine-Gordon model}
\label{sec:sGordon}

As a next example, let us discuss the sine-Gordon model, defined as
\begin{eqnarray}
{\cal A}^{\rm sG}  = \frac{c}{16 \pi} &&\int\limits_0^{1/T}\!d\tau\!\int\!dx  \nonumber \\
	&& \times \left[(\partial_x \Phi)^2 + \frac{1}{c^2}(\partial_\tau \Phi)^2 - g^2 {\rm cos}(\gamma' \Phi) \right]   \;.
\end{eqnarray}
Here $\Phi$ is the sine-Gordon field, $g$ and $\gamma'$ are coupling constants, and $c$ has the dimension of velocity. Diffusive behavior of the sine-Gordon model in the semiclassical limit was found first by Damle and Sachdev~\cite{sinegordon}. In this model the correlation function is defined by
\begin{equation}
C_\Phi (x,t)  = \langle e^{i\eta \Phi(x,t)} e^{-i\eta \Phi(0,0)}  \rangle_{{\cal A}^{\rm sG}} \;.\label{eq:def:sG-corr}
\end{equation}
Now we shall prove that the relaxation of correlations in equation~(\ref{eq:def:sG-corr}) is described by equation~(\ref{eq:relaxfv}), with the parameter $\gamma$ set to
\begin{equation} 
 \gamma=-{\rm cos} \left(\frac{2\pi \eta}{\gamma'}\right) \label{eq:sG-Pottsequiv} \;.
\end{equation}

\begin{figure}
	\centering
	\resizebox{0.95\columnwidth}{!}{\includegraphics[angle=270]{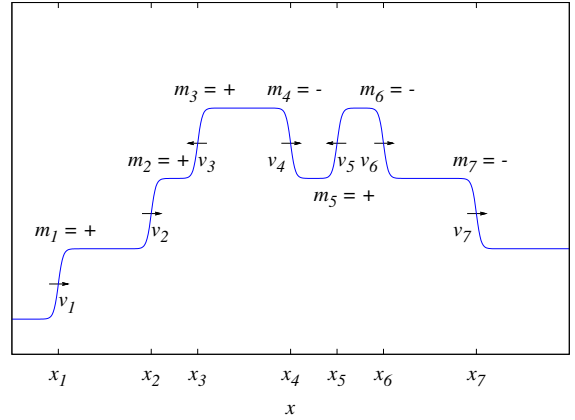}}
	\caption{ A configuration of kinks and antikinks in the sine-Gordon model at $t=0$. }
	\label{fig:kinks}
\end{figure}

Within the semiclassical approximation and using the results of the previous subsections, it is easy to show the equivalency of the correlators of the sine-Gordon model and the Potts model. Let us follow here a slightly different path than the one in reference~\cite{sinegordon}. It is known that low-energy configurations of the sine-Gordon field can be described in terms of domains, separated by soliton and anti-soliton trajectories.  These domains can be labeled by integers so that the field is approximately
\begin{equation}
\Phi(x,t) \approx \frac{2\pi}{\gamma'}\sum_\nu m_\nu \Theta(x-x_\nu(t))\;,
\end{equation}
where $x_\nu(t)$ are the trajectories of the solitons and $m_\nu =\pm1$ are the charges of the solitons (see Fig.~\ref{fig:kinks}). The collision of the solitons is described by a reflective \smatr~ \cite{sinegordon}.

Within the semiclassical approximation, we find that the average over the kinetic variables and soliton charges factorizes as
\begin{eqnarray}
C_\Phi(x,t)=& \sum\limits_n & \left\langle \delta_{n,\sum_\nu [\Theta(x - x_\nu - v_\nu t) -  \Theta(0 -x_\nu)]} \right\rangle_{\{x_\nu,v_\nu\}} \nonumber \\ 
& \times (-1)^{n} &\left\langle \sum_k P_k(n) e^{i\eta\frac{2\pi}{\gamma'}(2k-\vert n\vert)}\right\rangle_{\{m_\nu\}}. \label{eq:corrfunc_sinegordon}
\end{eqnarray}
Here $P_k(n)$ is the probability that the total number of $m_\nu = +1$-charge soliton lines crossing the section $(0,0) \to (x,t)$ is $k$, while the total number of collisions along the section is $n$~\cite{footnote1}. The second term in equation~(\ref{eq:corrfunc_sinegordon}) can be simplified as follows. The probabilities that a soliton or an anti-soliton is associated with a given trajectory are equal in a configuration. Therefore the probability for intersecting $k$ soliton lines has a binomial distribution
\begin{equation}
\left\langle P_k(n) \right\rangle_{\{m_\nu\}} ={\vert n \vert \choose k} \frac{1}{2^{\vert n\vert}}\;,  \label{eq:Binomial_sum}
\end{equation}
since every sequence of a given number of solitons and anti-solitons has equal probability. Using equation~(\ref{eq:Binomial_sum}), we can compute the sum over $k$ in equation~(\ref{eq:corrfunc_sinegordon}) to obtain the result
\begin{equation}
\left\langle \sum_k P_k(n) e^{i\eta\frac{2\pi}{\gamma'}(2k-\vert n\vert)}\right\rangle_{\{m_\nu\}} =  \cos{\left(\eta \frac{2\pi}{\gamma'}\right)}^{\vert n \vert}  \;.
\end{equation}
Comparing this expression with the relaxation function found for the Potts model one finds equation~(\ref{eq:sG-Pottsequiv}).

\section{Conclusions}
\label{sec:concl}

In this paper we obtained an analytical expression for the relaxation of certain correlation functions in a number of gapped one-dimensional systems, using the same semiclassical approach we applied earlier for the $Q$-state quantum Potts model~\cite{rapp-potts}. In particular, here we studied the $O(N)$ rotor models, discussed antiferromagnetic correlations in the $S=1$ spin Heisenberg model, and also analyzed the dynamical correlations of the sine-Gordon model.

Similar to the Potts model, all these systems have a gap in their quasiparticle spectra, and their quasiparticles posses some internal quantum numbers. In all these models, the collision of the quasiparticles is described by a reflective \smatr. The main result of our work is that the relaxation at finite temperature is given by the same universal function $R_\gamma(\bar x, \bar t)$, defined by equation~(\ref{eq:relaxfv}). Moreover, as shown in reference~\cite{rapp-potts}, the semiclassical relaxation function $R_\gamma(\bar x, \bar t)$ also describes relaxations in the transverse field Ising model, corresponding to $\gamma=1$.

For models with the parameter $\gamma < 1$ ( spin models having a multiplet of excitations), the expression derived here predicts that the long time behavior is diffusive. In the case of the $Q=3$ quantum Potts model on the ferromagnetic side, the origin of this diffusive behavior is the random diffusion of the domain 
containing the origin at time $t=0$, due to the random motion of the kinks which separate neighboring domains. On the other hand, in the case $\gamma \to 1$, the singularity at $\phi = \pi$ dominates the integral in equation~(\ref{eq:relaxfv}), leading to an exponential rather than diffusive decay. This exponential behavior  is a consequence of a "destructive interference ``: for the Ising model in a transverse field, the sequence of domains on the ferromagnetic side is always alternating. Carrying out the calculations assuming a non-interacting/transmissive \smatr~also leads to exponential decay in the correlation functions~\cite{rapp-potts}.

We thus find that the relaxation function equation~(\ref{eq:relaxfv}) is rather universal and it appears in 
essentially all one - dimensional models having a reflective \smatr. The idea of universal diffusive or exponential relaxations depending on the structure of the scattering matrix appeared already in references~\cite{rapp-potts,sinegordon}, and in reference~\cite{Konik-nuclphys}, although the results 
of the latter work do not entirely agree with the relaxation function obtained within the 
semiclassical approach. Apparently, the form-factor expansion of reference~\cite{Konik-nuclphys} builds in long-ranged correlations to the correlation function, since at $t=0$, it decays as $\sim \vert \bar x \vert ^{-1/2}$.  According to physical intuition, any spatial correlation between kink quantum numbers should decay beyond $\xi_c$. Therefore, on simple physical grounds, one expects exponential decrease in space for the equal time correlation function, which contradicts the result of reference~\cite{Konik-nuclphys}. The $\vert \bar t \vert^{-1/2}$ behavior for $x=0$, on the other hand, can be understood simply as a result of diffusive motion of the quasiparticles.~\cite{rapp-potts} The authors of reference~\cite{Konik-nuclphys}  have also shown that for a diagonal \smatr~ the decay becomes exponential. We believe, however,  that the generic \smatr~is of an exchange form. In fact, Damle and Sachdev argued in reference~\cite{sinegordon} that a transmissive \smatr~needs the fine tuning of an infinitely large number of coupling constants. A similar observation was made in the discussion of the \smatr~of the $Q=3$ state Potts model, where we argued that on a lattice, the scattering matrix is reflective in the low velocity limit.

We are thus tempted to call the relaxation function $R_\gamma(\bar x, \bar t)$ the universal semiclassical relaxation function for gapped spin systems. We must mention that the diffusive behavior discussed here derives from the asymptotic form of the \smatr, and it therefore has a somewhat limited  range of validity: At finite temperatures, the momenta of incoming particles remain finite, and therefore the scattering matrix is not purely reflective.  We estimate that the probability that a quantum number flips rather than being reflected during a collision is $P_{\rm flip} \sim a^2 \langle \Delta k^2 \rangle \sim a^2 T \Delta/c^2  $, where $a$ is the lattice spacing and $\Delta k$ is the momentum difference between the quasiparticles. These estimates carry over to all models considered here, since in all these models the \textit{amplitude} of non-reflective scattering in the leading order is proportional to the typical quasiparticle momentum.\cite{rapp-potts,sinegordon,semiclass-DamleSachdev,ON-smatrix,sGordon_smatrix} As a consequence, the diffusive decay must be asymptotically replaced by an exponential decay at very long time scales and very large separations. The timescale at which the diffusive behavior is expected to vanish is $t_{\rm diff} \sim \tau_c P_{\rm flip}^{-1} \sim (c^2 \Delta / a^2) e^{\Delta/T}$. The corresponding lengthscale is  $x_{\rm diff} \sim \sqrt{ D t_{\rm diff}} \sim c^2/(a T \Delta) e^{\Delta/T} $.

This research has  been supported by Hungarian grants OTKA Nos. NF061726, K73361, and NI70594, and by the SFB 608 of the DFG.



\begin{thebibliography}{99}
\bibitem{footnote1} Here we count trajectories coming from the left and right with opposite signs 
when we compute $k$ and $n$
\end{thebibliography}
\end{document}